\documentstyle[psfig]{mn2e}
\newif\ifAMStwofonts
\title[4C02.27: a quasar with episodic activity?]
      {4C02.27: a quasar with episodic activity?}
\author[M. Jamrozy et al.]
       {M. Jamrozy$^1$$\thanks{E-mail: jamrozy@oa.uj.edu.pl (MJ);
                   djs@ncra.tifr.res.in (DJS);
                   chiranjib.konar@gmail.com(CK)}$,  
        D.J. Saikia$^{2,3}$ and C. Konar$^4$ \\
$^1$ Obserwatorium Astronomiczne, Uniwersytet Jagiello\'nski, ul. Orla 171, PL-30244 Krak\'ow, Poland \\
$^2$ National Centre for Radio Astrophysics, TIFR, Ganeshkhind, Post Bag 3, Pune 411 007, India \\
$^3$ Australia Telescope National Facility, CSIRO, PO Box 76, Epping, NSW 1710, Australia \\
$^4$ Inter-University Centre for Astronomy and Astrophysics, Ganeshkhind, Post Bag 4, Pune 411 007, India \\
}
\date{Accepted.                         Received }

\pagerange{\pageref{firstpage}--\pageref{lastpage}}
\pubyear{  }

\begin{document}

\maketitle

\label{firstpage}

\begin{abstract}
Striking examples of episodic activity in active galactic nuclei 
are the double-double radio galaxies (DDRGs) with two pairs of oppositely-directed
radio lobes from two different cycles of activity. Although there are over
about a dozen good examples of DDRGs, so far no case of one associated 
with a quasar has been reported. We present Giant Metrewave Radio Telescope
observations of a candidate double-double radio quasar (DDRQ), J0935+0204 (4C02.27),  
and suggest that radio jets in this source may also have been 
intrinsically asymmetric, contributing to the large observed asymmetries
in the flux density and location of both pairs of radio lobes. 
\end{abstract}


\section{Introduction}
A striking example of episodic nuclear activity is when a new pair of radio lobes
is seen closer to the nucleus before the `old' and more distant radio
lobes have faded (e.g. Subrahmanyan, Saripalli \& Hunstead 1996; Lara et al. 1999).
Such sources have been christened as `double-double' radio galaxies
(DDRGs) by Schoenmakers et al. (2000a). 
More recently Brocksopp et al.  (2007) have also reported the discovery of 
a triple-double radio galaxy with three distinct cycles of activity. Although
more than approximately a dozen or so of such DDRGs are known in the literature (Saikia, 
Konar \& Kulkarni 2006, and references therein), so far no case of one associated
with a quasar has been reported. This could be partly due to difficulties in distinguishing
a knot of emission in a jet from a hotspot, since quasars tend to have more prominent
jets than galaxies due to effects of relativistic beaming. However, if radio galaxies
and quasars are intrinsically similar, one should find evidence of emission from an earlier    
cycle of activity in quasars as well. This may be in the form of distinct pairs of lobes as
in the DDRGs, or may be seen as diffuse relic emission beyond the extent of the younger
double lobed radio source, as seen for example in the radio galaxy 4C29.30 (Jamrozy et al. 2007). 

Since the lobes from an earlier cycle of activity are likely to have a steep radio spectrum,
low-frequency observations with telescopes such as the Giant Metrewave Radio Telescope (GMRT)
should help identify such features. However, although the DDRG J0041+3224 was discovered from
GMRT and Very Large Array (VLA) observations (Saikia et al. 2006), deep GMRT observations
of a few hundred sources at 153, 244, 610 and 1260 MHz have not yielded clear examples of 
sources with episodic activity (Sirothia et al. 2009).
We have been systematically making observations, searching the literature as well as 
images from surveys made with the VLA such as 
NVSS (NRAO VLA Sky Survey; Condon et al. 1998) and 
FIRST (Faint Images of the Radio Sky at Twenty-cm;
Becker, White \& Helfand 1995)  for candidate DDRGs and  DDRQs or quasars with signs
of episodic radio activity. Identifying a DDRG/DDRQ is not automated and each image
is examined by eye. In this paper, we present observations with the GMRT at 619 MHz 
of possible relic emission associated with the quasar, 4C02.27 (J0935+0204),
and discuss the possibility that this might be a quasar with signs of episodic 
activity. 

\begin{figure*}
\hbox{
 \psfig{file=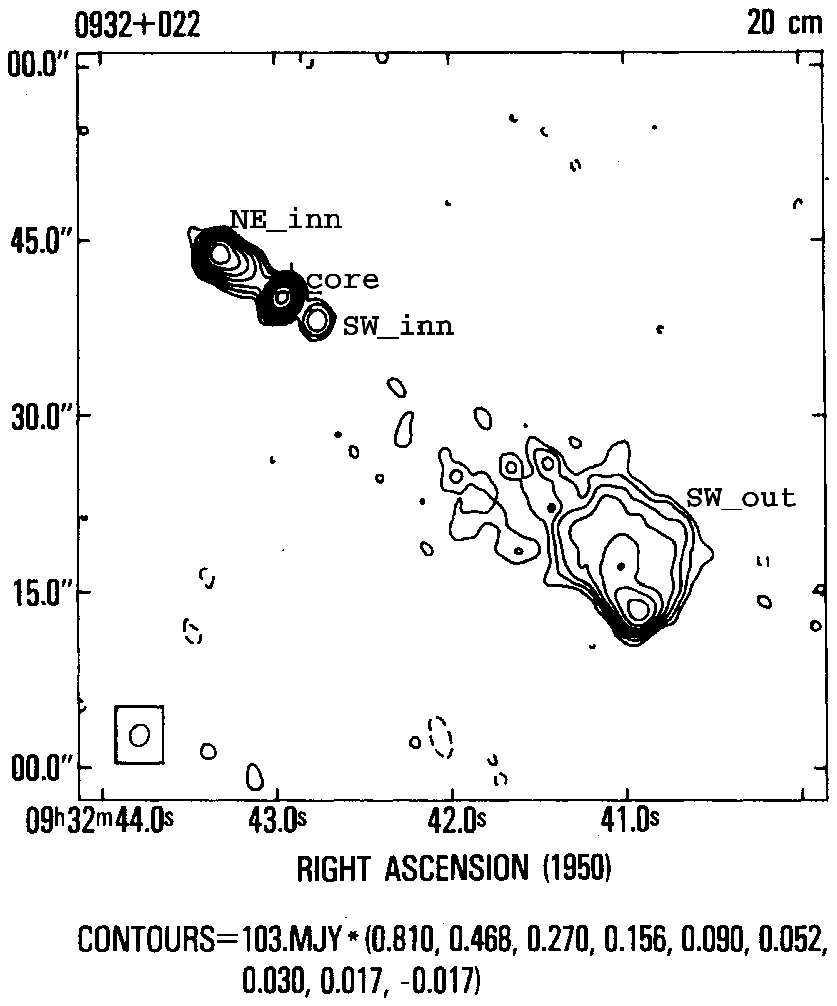,width=3.0in,angle=0}
 \psfig{file=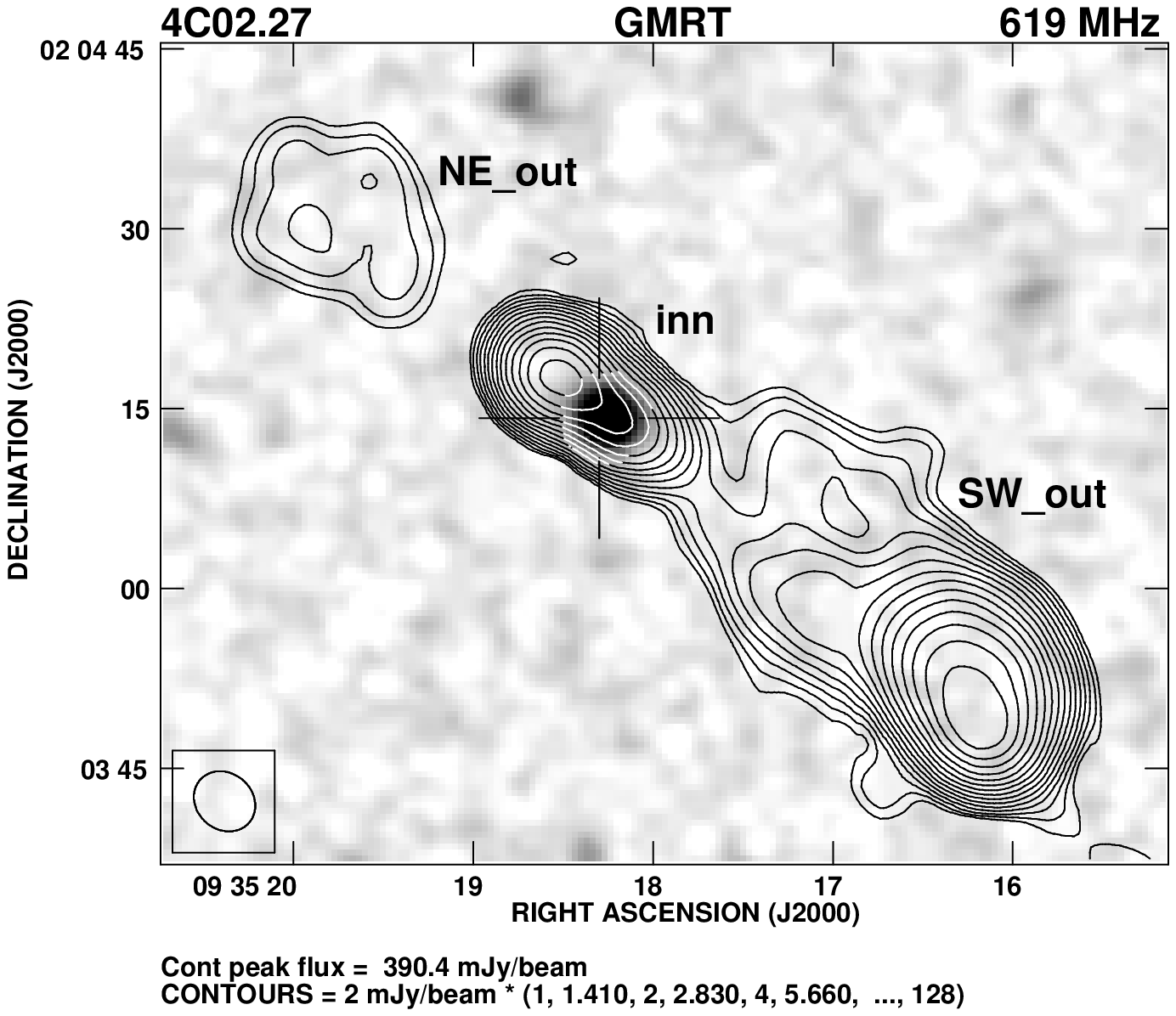,width=4.0in,angle=0}
}
\caption{Left panel: VLA image of 4C\,02.27 at 1413 MHz with an angular resolution of 
$\sim$1.7 arcsec reproduced from Hintzen et al. (1983), with the components NE$_{\rm inn}$,
SW$_{\rm inn}$ and SW$_{\rm out}$ marked. Right panel: GMRT image of 4C\,02.27 at 619 MHz
with an angular resolution of 5.42$\times$4.68 arcsec$^2$ along a PA of 49$^\circ$ showing
the outer north-eastern component, NE$_{\rm out}$. The cross marks the position of the 
quasar.
         }
\end{figure*}

\section{4C02.27}
The radio source 4C02.27 (J0935+0204) 
is associated with a quasar at a redshift of 0.649117$\pm$0.000210 as listed
in the NASA Extragalactic Database (NED) from measurements made as part of the
the SDSS (Sloan Digital Sky Survey; Abazajian et al. 2004;
Schneider et al. 2005). %\footnote{http://cas.sdss.org/dr7/en/tools/explore/obj.asp}).  
The corresponding luminosity distance is 3880.7 Mpc and 1 arcsec corresponds to 6.918 kpc 
in a Universe with 
H$_\circ$=71 km s$^{-1}$ Mpc$^{-1}$, $\Omega_m$=0.27, $\Omega_\Lambda$=0.73 (Spergel et al. 2003). 

The published radio images show a double-lobed radio source which is highly asymmetric in
both flux density and location of the outer components, and has two prominent
hotspots at the outer edges (labelled as SW$_{\rm out}$ and NE$_{\rm inn}$ in Fig. 1, left panel) with an 
overall angular separation of 47 arcsec (325 kpc). 
The separations of the peaks of emission in NE$_{\rm inn}$ and SW$_{\rm out}$ from the nucleus are
6.3 arcsec (43.6 kpc) and 40.6 arcsec (281 kpc) respectively (e.g. Hintzen, Ulvestad \& Owen 1983;
Swarup, Sinha \& Hilldrup 1984; Price et al. 1993). The corresponding arm-length ratio is
$\sim$6, making it one of the very asymmetric sources in terms of the location of the components.
The flux density ratio of the two components at 1400 MHz is 3.5, with SW$_{\rm out}$, which
is farther from the nucleus being also brighter. In addition, there is weak component (SW$_{\rm inn}$) 
with a flux density of 9 mJy at 1400 MHz located $\sim$3.5 arcsec (24.2 kpc)
south-west of the nucleus along the axis of the source. There is no evidence of a distinct 
jet-like structure as defined by Bridle \& Perley (1984). From VLA C-array observations at 
5 GHz, Saikia et al. (1984) find the south-western component to be 15.4 per cent polarised
compared with 5.7 per cent for the north-eastern one and 1.1 per cent for the core component.

In an optical study of the host galaxies of intermediate-redshift quasars, R\"onnback et al.
(1996) report an arm-like structure resembling a tidal tail in 4C02.27 and show that the
luminosity profile of the host galaxy follows an r$^{1/4}$ law. Their effective radius is 
consistent with earlier measurements by Romanishin \& Hintzen (1989) although R\"onnback et al. 
find the host galaxy magnitude to be $\sim$0.4 mag brighter.

In the next Section we present GMRT observations of the source at 619
MHz as well as the NVSS and FIRST images of the source.  In Section 3, we discuss the possibility
that 4C02.27 (J0935+0204) could be a DDRQ exhibiting signs of episodic activity.

\section{Outer north-eastern lobe}
In this section we present the results of the GMRT observations as well as the
NVSS and FIRST images. The GMRT and VLA images show the presence of a diffuse lobe beyond the 
north-eastern hotspot.

\subsection{GMRT observations and results}
The source was observed with the GMRT on 2007 September 01 at 619 MHz for 
approximately 330 minutes on source.  
The observations were made in the standard manner, with  each observation
of the target-source interspersed with observations of the phase calibrator,
0943$-$083.  3C286 and 3C147 were both observed  for flux density and bandpass 
calibration. The flux densities are on the Baars et al. (1977) scale.
The data collected were calibrated
and reduced in the standard way using the NRAO {\tt AIPS} software package.
Several rounds of self calibration were done to improve the quality of the images.
The GMRT image has an angular resolution of 5.42$\times$4.68 arcsec$^2$ along a 
position angle (PA) of 49$^\circ$ and an rms noise of 0.22 mJy beam$^{-1}$.  

The GMRT image (Figure 1, right panel) shows clearly a diffuse lobe of emission to the
north-east well beyond the eastern hotspot, which we will refer to as NE$_{\rm out}$. 
The total flux density of this feature is 
41 mJy, while the total flux density of the source is 1578 mJy. The total flux density
of the south-western lobe (SW$_{\rm out}$), is 1138 mJy, while that of the central feature which contains the
core and the inner north-eastern (NE$_{\rm inn}$) and south-western (SW$_{\rm inn}$) components 
is 393 mJy. 

\subsection{NVSS and FIRST images}
In Figure 2, we present the NVSS and FIRST images of the source at 1400 MHz with angular resolutions
of 45 and $\sim$6 arcsec respectively superimposed on the DSS image.  The FIRST image is very
similar to that of the GMRT one showing the diffuse outer lobe of emission to the
north east (NE$_{\rm out}$) in addition to the other features. The flux density of 
NE$_{\rm out}$ and SW$_{\rm out}$ are 15 mJy and 488 mJy, while that of the central 
feature consisting of the core and NE$_{\rm inn}$ and SW$_{\rm inn}$ is 230 mJy.

The spectral indices, $\alpha$ (S$\propto\nu^{-\alpha}$), of the outer features, 
NE$_{\rm out}$ and SW$_{\rm out}$, are $\sim$1.2 and 1.0 respectively between 610 and 1400 MHz, 
while the total flux densities yield a spectral index of $\sim$0.9. The two-point spectral 
index of the inner component consisting of the core, NE$_{\rm inn}$ and SW$_{\rm inn}$
is $\sim$0.7. The deconvolved size of the NVSS image, which is 52$\times$12 arcsec$^2$ along
a position angle of 52$^\circ$, is consistent with the structure of the source. The total
flux density of 787 mJy in the NVSS image is very similar to value of 800 mJy at 1413
MHz estimated by Hintzen et al. (1983). We have also examined the VLA Low-Frequency Sky 
Survey (VLSS; Cohen et al. 2007) image but this shows no new feature.

\begin{figure}
\hbox{
 \psfig{file=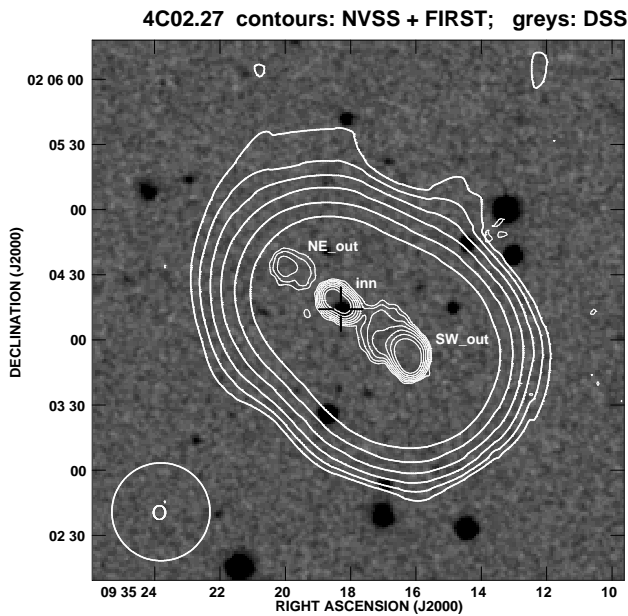,width=3.4in,angle=0}
}
\caption{1400-MHz VLA images of 4C\,02.27. NVSS and FIRST contour maps of the 
entire source overlayed on the optical field from the Digital Sky Survey (DSS). 
The contours are spaced by factors of 2 and the first contours are 1.35 and 0.6
mJy~beam$^{-1}$ respectively.  The sizes of the beams are indicated by ellipses in 
the bottom left corner of the image. The cross marks the position of the radio core.}
\end{figure}

\section{Episodic nature of 4C02.27}
The GMRT and the VLA FIRST images clearly show two pairs of components, the inner
and the outer doubles with angular sizes of $\sim$10 and 68 arcsec respectively, which
corresponds to projected linear sizes of $\sim$70 and 470 kpc respectively. The 
inner and outer lobes appear distinct and do not appear to be multiple 
hotspots in the lobes (cf. Lonsdale \& Barthel 1986). The 
eastern component of the inner double clearly shows an edge-brightened structure
similar to the hotspots of classical FRII sources. On the other hand, the western 
component of the inner double is compact and weak and its detailed structure is 
not well determined although the high-resolution image published by Price et al. 
(1993) hints that it too might have an edge-brightened structure. The flux density 
ratio of these two inner components is $\sim$17 at 1400 MHz while the peak 
brightness ratio is $\sim$6 (Hintzen et al. 1983). The inner double is  
reminiscent of the inner structure of the radio 
galaxy 3C219 which has been interpreted to be due a restarted jet activity (Bridle, 
Perley \& Henriksen 1986; Clarke et al. 1992), and has been included in samples
of double-double radio galaxies (Schoenmakers et al. 2000a; Saikia et al.
2006; Saripalli \& Mack 2007). Let us examine whether the asymmetry in location and brightness
of the components of the inner double in 4C02.27 can be understood in terms
of the relativistic beaming scenario.  Here the arm-length ratio, R$_\theta$ is given by
$(1+\beta {\rm cos}\phi)/(1-\beta {\rm cos}\phi)$ and the  brightness ratio is 
(R$_\theta$)$^{2+\alpha}$,
where v=$\beta$c is the component velocity and $\phi$ is the angle of inclination
of the jet axis to the line of sight. For a velocity of $\sim$0.4c and
$\phi\sim 40^\circ$, which is a reasonable value for a lobe-dominated quasar, and $\alpha\sim$1, 
R$_\theta$$\sim$1.9 and the brightness ratio is $\sim$6.7. These values are close
to the observed values of $\sim$1.8 and 6 estimated from the image of Hintzen et al. (1983),
suggesting consistency of the arm-length and peak brightness ratio with the 
relativistic beaming scenario with the NE$_{\rm inn}$ lobe approaching us. High-resolution 
depolarisation and rotation measure estimates would help to examine this possibility. 
The precise values of the brightness ratio at different locations will also be affected by 
the microphysics of the dissipation of energy by shocks along the flow of the jet. 
However, the flux density ratio is much larger with the NE$_{\rm inn}$ component being
stronger by $\sim$17, suggesting intrinsic asymmetries as well over the lifetime of the 
inner double.

The situation is even more complex for the outer double. From the GMRT 619-MHz image
the arm-length ratio of the outer lobes is $\sim$1.3, while the peak brightness ratio
is $\sim$60 and the total flux density ratio $\sim$30. The values estimated from the
FIRST image are similar and are in the opposite sense to that of the inner double, with
the SW$_{\rm out}$ lobe now being farther and also brighter. Considering the outer lobes
alone, it would be difficult to reconcile the arm-length and the large brightness ratios  
in the simple version of the relativistic beaming scenario. Environmental asymmetries
coupled with the effects of relativistic motion can produce a wide range of arm-length
and brightness asymmetries (e.g. Jeyakumar et al. 2005). For a lobe approaching us
and propagating through a dense medium, the oppositely directed lobes could be somewhat
symmetrically located with large asymmetries in brightness or flux density ratios.
In this case, the SW$_{\rm out}$ lobe would be approaching us, which would be
inconsistent with the inner double. On the other hand, if the NE$_{\rm out}$ lobe is
approaching us, one would expect it to be brighter due to the effects of relativistic
motion as well as a denser environment, inferred from its closer distance to the 
radio nucleus. These suggest that the radio jets may have been intrinsically asymmetric
during this cycle of activity. The possibility of intrinsically asymmetric jets
has been suggested earlier, for example, for the highly asymmetric double-lobed radio source 
B0500+630 (Saikia et al. 1996), weak-cored one-sided sources (Saikia et al. 1989, 1990),
the one-sided radio emission in 3C273 which has been imaged with a dynamic range of
10$^4$:1 (Davis, Muxlow \& Conway 1985; Conway \& Davis 1994), the inner jets of the 
radio galaxies M87 (Kovalev et al. 2007) and NGC6251 (Jones 1986; Jones \& Wehrle 2002), 
and the optical counter jet in the radio galaxy 3C66B (Fraix-Burnet 1997). 
On the theoretical side, Icke (1983) has suggested a hydrodynamical model for gas flows
in the nuclear region that can affect the de Laval nozzle causing one-sided jets, 
while Wang, Sulkanen \& Lovelace (1992) have explored models where the 
ratio of jet luminosities depends directly on the degree of asymmetry of the magnetic
field in the nuclear region. 

To explain the extreme asymmetries one might also consider the possibility that the 
orientation of the jet might change during the different cycles of activity; in the extreme
cases the approaching jet during one cycle may be the receding one in the next or vice versa.
However, the alignment of all the four components in 4C02.27 suggests that this is not a likely
scenario for this source.

Another interesting aspect on the episodic nature of 4C02.27 is the presence of a hotspot 
in the SW$_{\rm out}$ lobe. This is not unique to 4C02.27. While the outer doubles are often
diffuse as for example in J1453+3308 (Schoenmakers et al. 2000a; Konar et al. 2006), hotspots
are also sometimes seen in the outer lobes as for example in the northern lobe of B1834+620
(Schoenmakers et al. 2000b). This can be used to estimate the time scale of interruption of
energy supply.  For typical sizes of hotspots of a few to $\sim$10 kpc in large radio sources 
(cf. Jeyakumar \& Saikia 2000), the hotspots are expected to fade in about $\sim$10$^4$$-$10$^5$ yr 
(e.g. Kaiser, Schoenmakers \& R\"ottgering 2000). This is a small fraction of the time it takes
for the material in the jets of large radio sources to reach the hotspots from the radio core.
Therefore it is reasonable to assume that the hotspot fades soon after the last jet material 
passes through them. The presence of a hotspot in the SW$_{\rm out}$ lobe implies that it still 
receives jet material. The travel time of the jet material from the core to the hotspot, $t_{\rm j}$,
is given by $D_{\rm hs}$/v$_{\rm {jet}}$, where $D_{\rm hs}$ is the physical distance of the 
hotspot from the core and v$_{\rm {jet}}$ is the velocity of the jet. In our case, 
$t_{\rm j}$ is $\sim$2.9$\times$10$^6$ yr for an inclination angle, $\phi\sim40^\circ$, and a
jet velocity of 0.5c. However, this estimate will be affected by light travel time effects due
to the orientation of the source axis. For example, if the SW$_{\rm out}$ lobe is on the receding
side, it will take longer for the information to reach us compared with it being on the approaching
side. The observed time difference, $t_{\rm {obs}}$,  between the ejection of the last jet material 
and its arrival at the hotspot is $\sim$1.8 and 3.9 Myr depending on the orientation of the source. 
If the time scale of interruption of the jet is larger than $t_{\rm {obs}}$, the hotspot in the
SW$_{\rm out}$ lobe and the inner structure cannot be observed simultaneously. Therefore, the 
interruption of jet activity must be less than $t_{\rm {obs}}$. Also, within this time period
the inner double forms, so that the time scale for interruption is less than $t_{\rm {obs}}$.
It is also of interest to note that the time scale of interruption for this lobe with a hotspot
is much smaller than for say J1453+3308 which has diffuse outer lobes. The dynamical and 
spectral ages of the diffuse outer lobes of J1453+3308 are $\sim$215 and 50 Myr, while that of
the inner double is only $\sim$2 Myr, suggesting a much longer time scale of interruption
(Kaiser et al. 2000; Konar et al. 2006). 

%%%%%%%%%%%%%%%%%%%%%%%%%%%%%%%%%%%%%%%%%%%%%%%%%%%%%%%%%%%%%%%%%%%%%
\begin{table}
\caption{Large radio quasars}
\begin{tabular}{l c c c c l }
\hline
Source    & Alt.   & Red-   & LAS     & LLS     &  Refs \\
          & name   & shift  &         &         &             \\
          &        &(MHz)   &(arcsec) &(kpc)    &             \\
(1)       &  (2)   & (3)    & (4)     &  (5)    &    (6)      \\
\hline
J0439$-$2422 &                &  0.8400 & 128   &  979  & 1    \\
J0631$-$5405 &                &  0.2036 & 312   & 1034  & 2    \\
J0750$+$6541 &                &  0.7470 & 222   & 1627  & 3      \\
J0810$-$6800 &                &  0.2311 & 390   & 1425  & 2     \\
J1027$-$2312 &                &  0.3090 & 198   &  893  & 1    \\
J1130$-$1320 &                &  0.6337 & 297   & 2033  & 4    \\
J1353$+$2631 &                &  0.3100 & 190   &  859  & 5    \\
J1427$+$2632 &                &  0.3660 & 240   & 1212  & 5      \\
J1432$+$1548 &                &  1.0050 & 168   & 1353  & 6    \\
J1504$+$6856 &                &  0.3180 & 204   &  939  & 3      \\
J1723$+$3417 & 4C34.47        &  0.2060 & 244   &  816  & 7,8    \\
J2042$+$7508 & 4C74.26        &  0.1040 & 610   &  1151 & 9      \\
\hline
\end{tabular}

1: Ishwara-Chandra \& Saikia (1999); 
2: Saripalli et al. (2005); 
3: Lara et al. (2001); 
4: Bhatnagar, Gopal-Krishna \& Wisotzki (1998);
5: Rogora, Padrielli \& de Ruiter (1986)
6: Singal, Konar \& Saikia (2004);
7: J\"agers et al. (1982); 8: Hooimeyer et al. (1992); 
9: Riley et al. (1988)

\end{table}
%%%%%%%%%%%%%%%%%%%%%%%%%%%%%%%%%%%%%%%%%%%%%%%%%%%%%%%%%%%%%%%%%%%%%

\section{Concluding remarks}
The quasar 4C02.27 with an overall linear size of $\sim$470 kpc appears to exhibit
signs of episodic activity and we classify it as a DDRQ. The source also exhibits
evidence of an intrinsic asymmetry of the oppositely-directed jets. Although most DDRGs appear to be
associated with large radio galaxies, often with sizes larger than approximately 1 Mpc,
signs of episodic activity are seen in smaller sources as well (Schoenmakers et al. 2000a,b;
Saikia et al. 2006 and references therein). Assuming an inclination angle of 40$^\circ$
to the line of sight, the intrinsic size of 4C02.27 would be $\sim$730 kpc, comparable
to some of the known or candidate DDRGs (Jamrozy et al. 2009). Considering the tendency
for DDRGs to often have large sizes, we examined the structures of large quasars with
sizes larger than $\sim$800 kpc (Table 1), including the quasar type object J0750$+$6541
(Lara et al. 2001). These do not show evidence of episodic activity,
which along with our search of the literature suggests that such objects are not common.
A deep low-frequency search amongst 374 sources, most of which were small, did not
show any clear example of a DDRG/DDRQ (Sirothia et al. 2009). Considering only giant radio galaxies, 
there are 4 DDRGs in the well defined sample of 49 sources (Schoenmakers 1999), suggesting 
that $\sim$10 per cent of these large sources may show signs of episodic activity. Considering 
the known giant radio galaxies selected from different samples and sometimes with incomplete 
structural information (e.g. Ishwara-Chandra \& Saikia 1999; Lara et al. 2001; Machalski, Jamrozy
\& Zola 2001; Saripalli et al. 2005), also suggests a similar percentage. Although the numbers
are very small at present, a similar fraction would be consistent with the unified scheme for
radio galaxies and quasars (e.g. Barthel 1989). 

\section*{Acknowledgments} 
We thank an anonymous reviewer and the editor for their comments which have significantly
improved the paper, and the staff of GMRT for their help with the observations. 
MJ acknowledges the MNiSW funds for scientific research during the years
2009--2012 under contract No 3812/B/H03/2009/36.
The GMRT is a national facility operated by the NCRA, TIFR. 
The National Radio Astronomy Observatory  is a facility of the National Science Foundation
operated under co-operative agreement by Associated Universities Inc. 
This research has made use of the NASA/IPAC extragalactic database (NED) which is operated
by the Jet Propulsion Laboratory, Caltech, under contract with the National Aeronautics
and Space Administration. We acknowledge use of the  Digitized Sky Surveys which were produced 
at the Space Telescope Science Institute under U.S. Government grant NAG W-2166. The images of 
these surveys are based on photographic data obtained using the Oschin Schmidt Telescope on Palomar 
Mountain and the UK Schmidt Telescope.  

{}

\end{document}